\documentclass[superscriptaddress,twocolumn,showpacs,preprintnumbers,prl]{revtex4-1}
\usepackage{amsbsy,amssymb,amsmath,bm}
\usepackage{graphicx,color,epsfig}
\usepackage{txfonts,xspace}
\usepackage{fancyhdr}
\usepackage[usenames,dvipsnames,svgnames,table]{xcolor}
\usepackage{hyperref}
\usepackage{ulem}

\def\bbbc{{\mathchoice {\setbox0=\hbox{$\displaystyle\rm C$}\hbox{\hbox
to0pt{\kern0.4\wd0\vrule height0.9\ht0\hss}\box0}}
{\setbox0=\hbox{$\textstyle\rm C$}\hbox{\hbox
to0pt{\kern0.4\wd0\vrule height0.9\ht0\hss}\box0}}
{\setbox0=\hbox{$\scriptstyle\rm C$}\hbox{\hbox
to0pt{\kern0.4\wd0\vrule height0.9\ht0\hss}\box0}}
{\setbox0=\hbox{$\scriptscriptstyle\rm C$}\hbox{\hbox
to0pt{\kern0.4\wd0\vrule height0.9\ht0\hss}\box0}}}}

\newcommand{\iu}{{i\mkern1mu}}

\newcommand{\ignore}[1]{}
\newcommand{\mComment}[1]{}
\newcommand{\gComment}[1]{}
\newcommand{\jComment}[1]{}
\newcommand{\rComment}[1]{}
\newcommand{\lComment}[1]{}

\newcommand{\Hsat}{H_\text{sat}}
\renewcommand{\gComment}[1]{\textcolor{magenta}{Gerardo: #1}}
\begin{document}
\title{%
  Three-Dimensional Crystallization of Vortex Strings in Frustrated Quantum Magnets
}
\author{Zhentao Wang}
\affiliation{Department of Physics and Astronomy, Rice University, Houston, Texas 77005, USA}
\author{Yoshitomo Kamiya}
\affiliation{iTHES Research Group and Condensed Matter Theory Laboratory, RIKEN, Wako, Saitama 351-0198, Japan}
\author{Andriy H. Nevidomskyy}
\affiliation{Department of Physics and Astronomy, Rice University, Houston, Texas 77005, USA}
\author{Cristian D. Batista}
\affiliation{Theoretical Division, T-4 and CNLS, Los Alamos National Laboratory, Los Alamos, New Mexico 87545, USA}

\begin{abstract}
  We demonstrate that frustrated exchange interactions can produce exotic 3D crystals of vortex strings
  near the saturation  field ($H=\Hsat$) of body- and face-centered cubic Mott insulators. The combination of cubic symmetry and  frustration leads to a
magnon spectrum of the fully polarized spin state ($H>\Hsat$) with degenerate minima at multiple {\it noncoplanar} $\bm{\mathit Q}$ vectors. This spectrum becomes gapless at  
the quantum critical point $H=\Hsat$ 
and the  magnetic ordering  below $\Hsat$ can be formally described as a condensate of a dilute gas of bosons. By expanding in the lattice gas parameter, we find that different vortex crystals span sizable regions of the phase diagrams for isotropic exchange and are further stabilized by symmetric exchange anisotropy.
\end{abstract}

\maketitle

Topological spin structures are of great potential  interest in future applications of spin-electronic techniques~\cite{Wolf2001}. The skyrmion crystals   discovered in the \textit{B}20-structure metallic alloys MnSi and Fe${}_{1-x}$Co${}_x$~\cite{Muhlbauer2009,Munzer2010,Yu2010} and in the Mott insulator Cu${}_2$OSeO${}_3$~\cite{Seki2012,Adams2012} are prominent examples. 
While the  emergence of  crystals of topological structures is reminiscent of the Abrikosov vortex lattice of type-II superconductors~\cite{Abrikosov1957,Essmann1967}, their origin is completely different in magnets. The basic difference 
is that magnetic systems are neutral Bose gases~\cite{Zapf2014}, while the charged Cooper pairs  are coupled to the electromagnetic gauge field. In other words,
the orbital coupling to an external field that stabilizes the Abrikosov vortex crystal in type-II superconductors is basically absent  in magnets. 

Topological spin structures must then be stabilized by other means. A key aspect of magnetic systems is that 
competing interactions are ubiquitous. A common outcome of this competition is a magnetic susceptibility that is maximized by several low-symmetry wave vectors $\bm{\mathit Q}$ connected by point-group transformations of the underlying material. Topological spin structures can emerge when the effective interaction between the different $\bm{\mathit Q}$ modes favors a  multi-$\bm{\mathit Q}$ ordering. This is the case of the \textit{B}20 materials, in which the Dzyaloshinskii-Moriya~\cite{Dzyaloshinsky1958,Moriya1960} interaction $D$ that arises from their noncentrosymmetric nature shifts the susceptibility maximum from $\bm{\mathit Q}={\bm 0}$ favored by the ferromagnetic exchange $J$ to a finite vector $|\bm{\mathit Q}| \simeq D/J$ that can have different orientations due to the cubic symmetry of the \textit{B}20 structure. Thermal fluctuations then play an important role for stabilizing the 6-$\bm{\mathit Q}$ structure that leads to the hexagonal skyrmion crystals in bulk versions of the \textit{B}20 materials~\cite{Muhlbauer2009}. In contrast, the phase is already stable at the mean field level in 2D thin films~\cite{Seki2012}. In addition to chiral magnets,
skyrmion crystals~\cite{Okubo2012}, soliton crystals~\cite{Kamiya2012,Selke1988}, and $Z_2$ vortex crystals~\cite{Rousochatzakis2012} have been theoretically predicted in other classical spin systems. 
All of these examples correspond to 2D crystals of topological structures, i.e., they are not modulated along the third dimension.

More recently, two of us  proposed the realization of magnetic vortex crystals in a quantum spin system of weakly coupled triangular layers near a magnetic field-induced quantum critical point (QCP)~\cite{Kamiya2014}. 
The basic idea is to use {\it geometric frustration} as the source of competing interactions and {\it quantum fluctuations} to stabilize the multi-$\bm{\mathit Q}$ vortex crystal states. This study focuses on a case with six degenerate {\it coplanar} $\bm{\mathit Q}$ vectors that are connected by the $C_6$ symmetry transformations of the underlying lattice. Consequently, as in the previous examples, the resulting vortex crystal is not modulated along the third direction.

In this Letter, we demonstrate that a similar mechanism can also stabilize  exotic 3D  crystals of vortex lines. Unlike the case of the 2D vortex crystals, we are unaware of alternative realizations of 3D vortex crystals. As we explained above, the observation of magnetic skyrmion lattices unveiled the relevance of multi-$\bm{\mathit Q}$ orderings that produce 2D crystals of topological structures. However, much less effort has been devoted to  the 3D crystals that can also arise from multi-$\bm{\mathit Q}$ orderings. The recent real-space observation of a skyrmion-antiskyrmion cubic lattice in MnGe~\cite{Tanigaki2015} confirms the physical relevance of these 3D structures. The key to realize 3D crystals of topological objects is to find regions of stability of multiple {\it noncoplanar}-$\bm{\mathit Q}$ orderings. Consequently, we study the body-centered (bcc) and face-centered cubic (fcc) lattices that commonly occur in nature (typical examples are the transition-metal oxides and fluorides~\cite{Goodenough1963}, solid ${}^3$He~\cite{Roger1983}, 3D Wigner crystals~\cite{Candido2004}, and the alkali-metal fulleride Cs${}_3$C${}_{60}$~\cite{Takabayashi2009,Ganin2010}). 
By extending the exchange interactions up to third nearest neighbors, we produce a single-magnon dispersion with multiple degenerate minima at noncoplanar $\bm{\mathit Q}$ vectors connected by the cubic point group. We compute the optimal single-particle state for condensing  the magnons and find that several multi-$\bm{\mathit Q}$ states corresponding to different  vortex crystals span sizable regions of the phase diagrams with isotropic exchange. These phases are further stabilized by symmetric exchange anisotropy that arises from, e.g., dipole-dipole interactions or spin-orbit coupling. The resulting spin textures consist of exotic 3D patterns of vortex strings. 


We consider a spin-$\tfrac{1}{2}$ Heisenberg model on bcc and fcc lattices coupled to a magnetic field:
\begin{equation}\label{Eq:Hamiltonian}
\hat{H}=\sum_{\langle ij \rangle} J_{ij} \bm{S}_i \cdot \bm{S}_j - \sum_i \bm{H} \cdot \bm{S_i},
\end{equation}
where $J_{ij}$ are the Heisenberg interactions up to 3rd nearest neighbor \{$J_1,J_2,J_3$\}. In this study, we focus on the external field $\bm{H}$ applied along the high symmetry [111] direction. The spin-$\tfrac{1}{2}$ operators can be represented by hard-core bosons~\cite{Matsubara1956}: $S_i^+=b_i,\,S_i^-=b_i^\dagger,\, S_i^z=1/2-b_i^\dagger b_i$, where the $z$ axis is along the magnetic field direction. The Hamiltonian is thus transformed into a model for an interacting Bose gas:
\begin{equation}\label{Eq:BH}
\hat{H} \!\!=\!\!\sum_{\bm{k}}\! ( \omega_{\bm{k}}-\mu ) b_{\bm{k}}^\dagger b_{\bm{k}}+\frac{1}{2N} \!\!\sum_{\bm{k},\bm{k}^\prime,\bm{q}} \!\!(U+V_{\bm{q}}) b_{\bm{k}+\bm{q}}^\dagger b_{\bm{k}^\prime - \bm{q}}^\dagger b_{\bm{k}^\prime} b_{\bm{k}},
\end{equation}
where $\omega_{\bm{k}}$ is the single-boson (magnon) dispersion, $\mu=\Hsat-H$ is the chemical potential, $U \to \infty$ is the on-site hard-core potential, and $V_{\bm{q}}$ (Fourier transform of $J_{ij}$) is the density-density interaction arising from the Ising component of the spin exchange~\cite{supp}. 

\begin{figure}[!tbp]
  \includegraphics[width=0.4\textwidth]{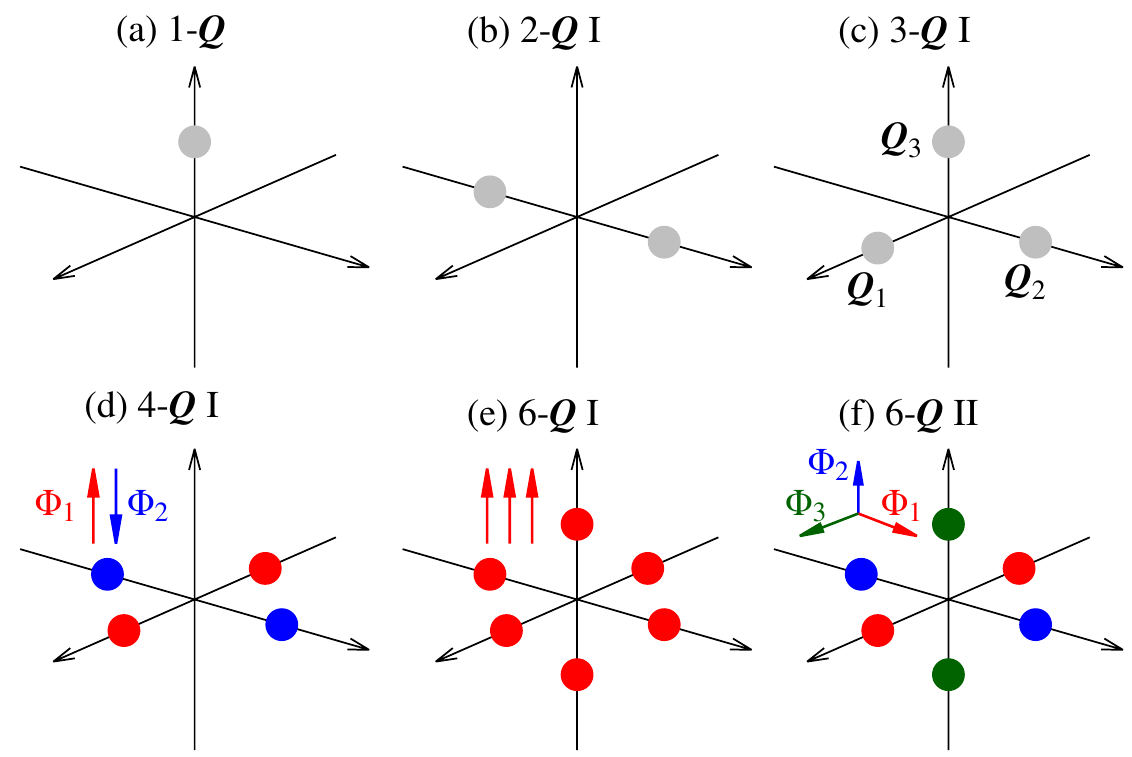}
  \caption{(color online).
    \label{Fig:states_label}
    (a)-(f) Schematic momentum-space representations of the multi-$\bm{\mathit Q}$ condensates near the field-induced QCP for  the case of six degenerate minima. The arrows representing the phases $\Phi_n$ of the $\bm{\mathit Q}_n$ component of the order parameter [see  Eq.~\eqref{Eq:energy}] are only shown for states in which their relative values are fixed by the interactions or anisotropy. The gray (light) color indicates no correlation among the different phases 
$\Phi_n$.}
  \vspace{-6pt}
\end{figure}

\begin{figure}[!tbp]
  \includegraphics[width=0.48\textwidth]{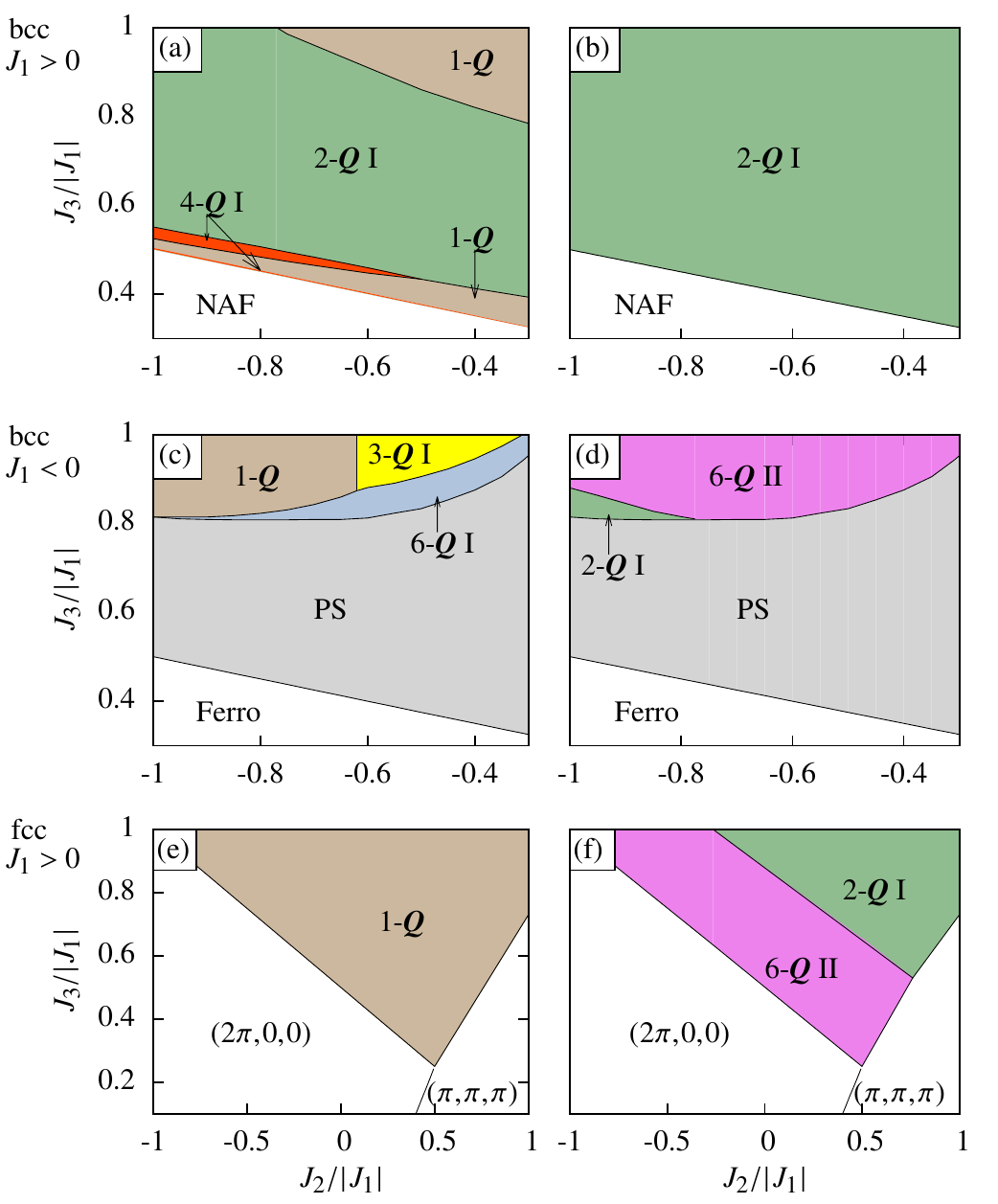}
  \caption{(color online).
    \label{Fig:phd}
    Phase diagrams of the Heisenberg model \eqref{Eq:Hamiltonian} under the nearly saturated magnetic field, where $\omega_{\bm{k}}$ has six degenerate minima in the colored phases. (a) bcc lattice, $J_1>0$, no anisotropy.
    ``NAF'' denotes the case with $\bm{\mathit Q} = (2\pi,2\pi,2\pi)$. (b) bcc lattice, $J_1>0$, anisotropy dominating region. (c) bcc lattice, $J_1<0$, without anisotropy. ``Ferro'' denotes $\bm{\mathit Q}$ at $(0,0,0)$, and ``PS'' denotes regions where we have phase separation or bound states. (d) bcc lattice, $J_1<0$, anisotropy dominating region. (e) fcc lattice, $J_1>0$, no anisotropy. (f) fcc lattice, anisotropy dominating region.}
  \vspace{-6pt}
\end{figure}

The  relative strengths of  $J_1$, $J_2$, and $J_3$ determine the number of degenerate minima in the single-magnon dispersion $\omega_{\bm{k}}$. Phases with six and eight minima exist in both the bcc and fcc lattices. A phase with twelve minima also exists in bcc lattice~\cite{supp}. For concreteness, we will focus  on the region with six degenerate minima, whose positions are denoted by $\pm \bm{\mathit Q}_n=\pm Q \,{\bf \hat{e}}_n$, where $n=1,2,3$. 

The single-magnon dispersion becomes gapless at  $H=\Hsat$ which signals the phase transition into a Bose-Einstein condensate~\cite{Feldman1990,Feldman1991,Nikuni1995,Kotov1998,Giamarchi1999,Nikuni2000,Ueda2009,Marmorini2014,Kamiya2014}. In the vicinity of this transition $|H| \lesssim |\Hsat|$, the boson density is vanishingly small, and we can use Beliaev's dilute boson approach~\cite{Beliaev1958} to compute the effective boson-boson interactions in the long-wavelength limit. Because this is a controlled expansion in the small lattice gas parameter (ratio between the scattering length and the average interparticle distance), the result is asymptotically exact in the dilute limit. The next step is to condense the bosons in the most general single-particle state, i.e., to replace the bosonic operators for each wave vector $\bm{\mathit Q}$ by six complex amplitudes: $\langle b_{\pm \bm{\mathit Q}_n} \rangle/\sqrt{N}= \sqrt{\rho_{\pm \bm{\bm{\mathit Q}_n}}} \exp \left( i \phi_{\pm \bm{\mathit Q}_n}\right)$. The total energy is the sum of the low-energy terms  allowed by translation symmetry, i.e., density-density interactions between bosons in the same ($\Gamma_1$) and  different ($\Gamma_2, \Gamma_3$) minima, as well as a $\Gamma_4$ vertex that scatters  bosons between two pairs of opposite minima,
\begin{eqnarray}\label{Eq:energy}
  E \!&=& \! \frac{\Gamma_1}{2} \!\!\! \sum_{n, \sigma=\pm} \!\!\!  \rho_{\sigma \bm{\mathit Q}_n}^2 
  \!+ \Gamma_2 \sum_{n} \rho_{\bm{\mathit Q}_n}^{\;} \rho_{-\bm{\mathit Q}_n}^{}
  \!+ \Gamma_3  \!\!\!\!\!  \sum_{\substack{{n<m}\\ \sigma_1,\sigma_2=\pm}} \!\!\! \rho_{\sigma_{\!1}^{}\bm{\mathit Q}_n}^{}\rho_{\sigma_{\!2}^{}\bm{\mathit Q}_m}^{}
  \nonumber \\
  &+& \! 2\Gamma_4 \! \sum_{n<m} \!\! \sqrt{\rho_{\bm{\mathit Q}_n}^{}\rho_{-\bm{\mathit Q}_n}^{} \rho_{\bm{\mathit Q}_m}^{}\rho_{-\bm{\mathit Q}_m}^{}} \cos \left( \Phi_n\! - \! \Phi_m \right) -\mu \rho,
\end{eqnarray}
where $\rho = \sum_n \left( \rho_{\bm{\mathit Q}_n}+\rho_{-\bm{\mathit Q}_n} \right)$ is the total boson density and $\Phi_n=\phi_{\bm{\mathit Q}_n}+\phi_{-\bm{\mathit Q}_n}$. The interaction vertices $\Gamma_1,\ldots, \Gamma_4$ are obtained by summing over the ladder diagrams at zero total frequency~\cite{supp}.

The zero-temperature phase diagram is determined by minimizing the total energy $E$ given in Eq.~\eqref{Eq:energy}\cite{mathematica}. Depending on the relative strengths of exchange interactions, one of the six possibilities in Fig.~\ref{Fig:states_label} is realized.  Out of these, three condensates in particular realize vortex crystals: 3-$\bm{\mathit Q}$ I, 4-$\bm{\mathit Q}$ I, and 6-$\bm{\mathit Q}$ II (Fig.~\ref{Fig:states_label}). Another reason for considering these states is that the latter two are further stabilized by symmetric exchange anisotropy originated from spin-orbit coupling or dipole-dipole interactions. Close to the saturation field $\Hsat$, this exchange anisotropy yields the interaction term~\cite{supp}:
\begin{equation}
E_A \propto J_A \sum_n \sqrt{\rho_{\bm{\mathit Q}_n} \rho_{-\bm{\mathit Q}_n}} \cos (\Phi_n +2n \pi/3-\pi/2).
\label{ean}
\end{equation}
Although $J_A$ is typically small, $E_A \sim |\rho|$ is linear in the boson density. Consequently, in the dilute limit ($\rho \ll |J_A/J_1| \ll 1$), it always dominates over the exchange interaction in Eq.~\eqref{Eq:energy}. The calculations show that the 2-$\bm{\mathit Q}$ I, 4-$\bm{\mathit Q}$ I, and 6-$\bm{\mathit Q}$ II condensates are the three lowest energy states, degenerate to linear order in the density $\rho$. The degeneracy is lifted by further considering the second-order density-density interactions in Eq.~\eqref{Eq:energy}, stabilizing the 6-$\bm{\mathit Q}$ II state over a wide range of parameters on both bcc and fcc lattices (see Fig.~\ref{Fig:phd}). On the other hand, sufficiently far away from the QCP (but still in the low density regime, $|J_A/J_1| \ll \rho \ll 1$), $E_A$ is negligible and Eq.~\eqref{Eq:energy} alone determines the ground state configuration. 
The resulting phase diagrams for negligible (dominant) anisotropy are shown in the left (right) column of Fig.~\ref{Fig:phd}, respectively.  

\begin{figure}[!tbp]
  \includegraphics[width=0.41\textwidth]{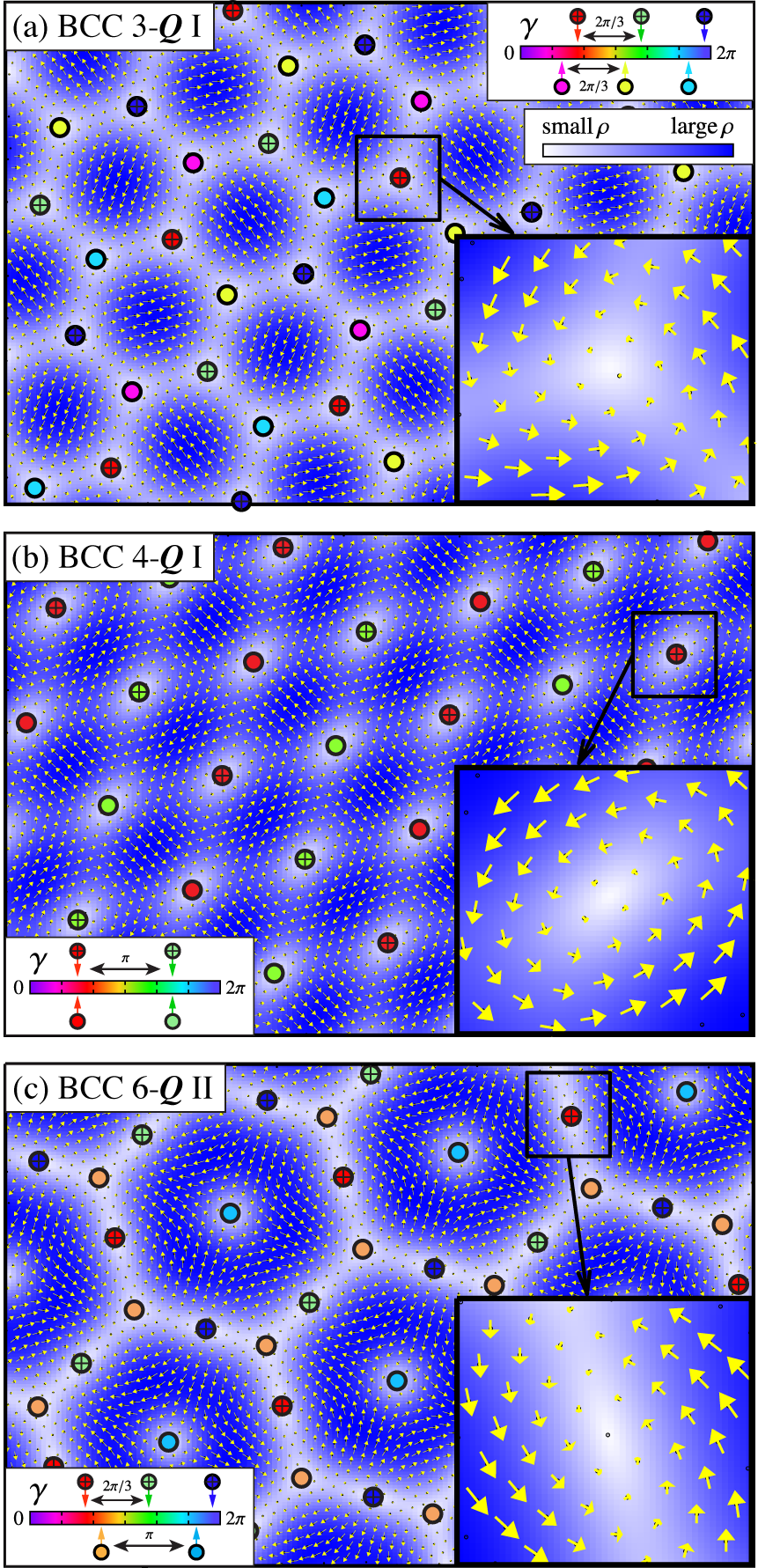}
  \caption{(color online). %
    \label{Fig:spin}
    Spin structures in the vortex crystal phases on [111] layers (${\bf H}  \parallel [111]$); only the $xy$ components are plotted.
    The color intensity denotes the boson density $\rho$ (the transverse spin components are $\sim \!\!\!\sqrt{\rho}$) and the bright spots denote fully polarized spins.
    The circles with (without) crosses denote vortex (antivortex) cores.
    Different colors of the circles denote different helicities $\gamma$ (see text).
    (a) The 3-$\bm{\mathit Q}$ I state on the bcc lattice for $|\bm{\mathit Q}_n| \ll 1$. (b) The 4-$\bm{\mathit Q}$ I state on the bcc lattice for $|\bm{\mathit Q}_n| \lesssim 2\pi$. (c) The 6-$\bm{\mathit Q}$ II state on the bcc lattice for $|\bm{\mathit Q}_n| \ll 1$.
  }
  \vspace{-6pt}
\end{figure}

\begin{figure}[!tbp]
  \includegraphics[width=0.47\textwidth]{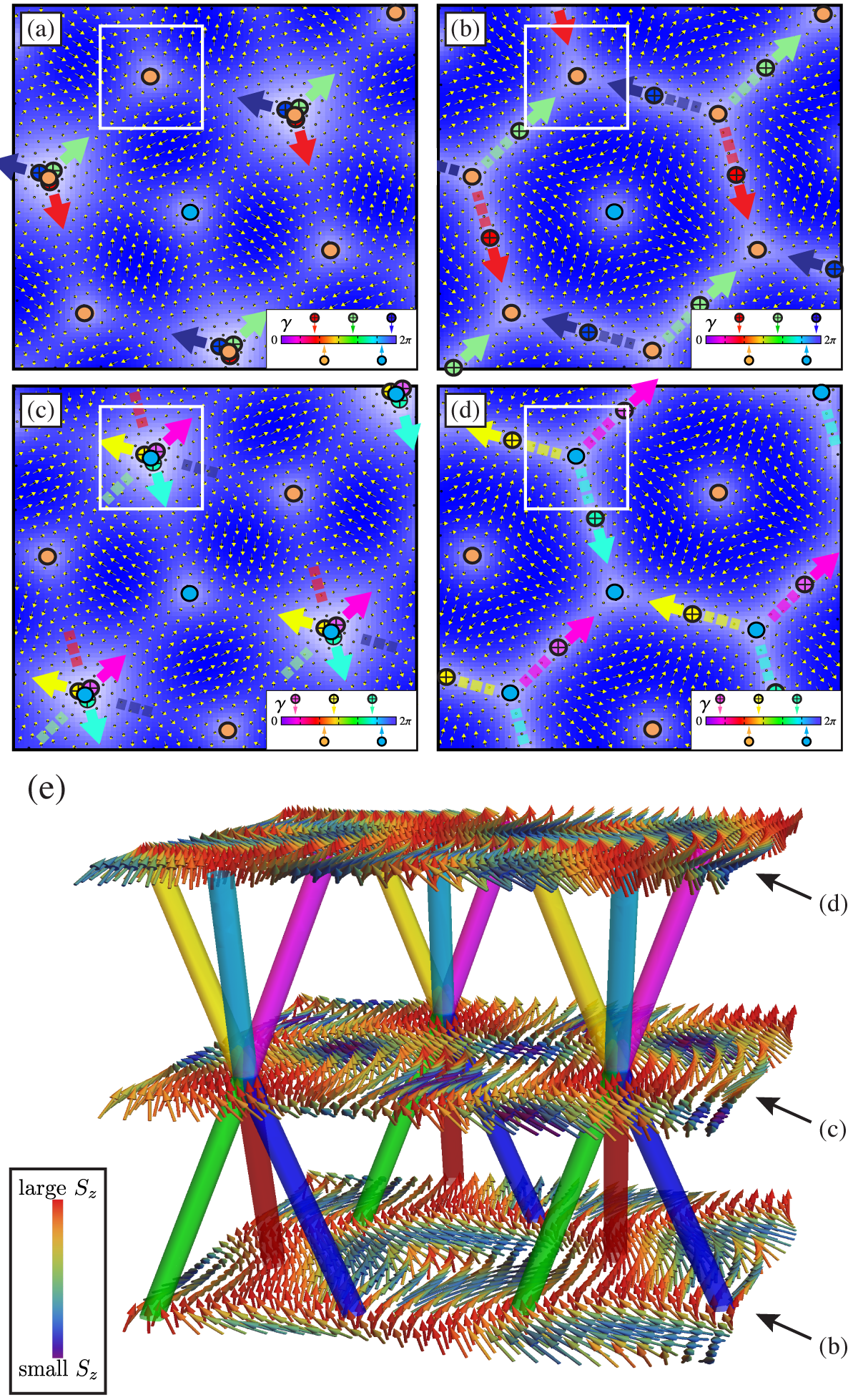}
  \caption{(color).
    \label{Fig:vortex_string}
    (a)-(d)
    The sequence of layers along the [111] direction for the 6-$\bm{\mathit Q}$ II state on the bcc lattice in the case $|\bm{\mathit Q}_n| \ll 1$. Several intervening layers between (a) and (b), etc., are omitted for simplicity. The arrows indicate how the vortex cores (circles with crosses) move from one layer to the next layer above. The lines of the antivortex cores (circles without crosses) are parallel to the [111] direction. The region enclosed by a square follows one of the antivortex core. The $\pi$ helicity shift occurs below the layer (c).
    (e) Three-dimensional picture of the vortex string structure. The strings corresponding to antivortices in (b)-(d) are not shown.
  } 
  \vspace{-6pt}
\end{figure}

We now focus on the 3-$\bm{\mathit Q}$ I, 4-$\bm{\mathit Q}$ I, and 6-$\bm{\mathit Q}$ II states that realize vortex crystals whose spin structures on [111] layers are illustrated in Fig.~\ref{Fig:spin}. We find that the vortex and antivortex cores form regular lattices on every layer. Vortices and antivortices correspond to different signs $\kappa = \pm1$ of the vector spin chirality $\bm{S}_{i} \times \bm{S}_{j}$ when we circulate ($i \to j$) around the vortex core. Explicitly, in the vicinity of the (anti)vortex core $\arctan (S_i^y/S_i^x)\!=\!\arg \langle{b_i^{\dagger}}\rangle \! \sim \! \kappa \varphi + \gamma + \delta(\varphi)$, where $\gamma$ is the helicity~\cite{Nagaosa2013} and $\delta(\varphi)$ is a $2\pi$-periodic function of the polar angle $\varphi$ around the (anti)vortex core in the [111] plane such that $\int^{2 \pi}_0 d \varphi \, \delta(\varphi)=0 $  and $\lvert{\delta(\varphi)}\rvert \ll 2\pi$. For a given chirality, the (anti)vortices can have different relative helicities, as we show in Fig.~\ref{Fig:spin}: the 3-$\bm{\mathit Q}$~I state includes three types of vortices with helicities that differ by $2\pi/3$ in each [111] plane (the same is also true for the  antivortices), see Fig.~\ref{Fig:spin}a; similarly, two types of (anti)vortices appear in the 4-$\bm{\mathit Q}$ I state with helicities that differ by $\pi$ [Fig.~\ref{Fig:spin}b]; the 6-$\bm{\mathit Q}$ II state contains three types of vortices with helicities that differ by $2\pi/3$ and two types of antivortices with helicities that differ by $\pi$ [Fig.~\ref{Fig:spin}c].  

The vortex cores are strings that extend along the third dimension. These strings form different patterns for each multi-$\bm{\mathit Q}$ condensate. The vortex and antivortex strings form parallel straight lines along the [111] direction  in the 3-$\bm{\mathit Q}$ I and 4-$\bm{\mathit Q}$ I states. The same is true for the antivortices of the 6-$\bm{\mathit Q}$ II states. However,  the vortex strings form a more exotic pattern in the 6-$\bm{\mathit Q}$ II state. As is shown in Fig.~\ref{Fig:vortex_string},  the vortex strings cross each other and the helicities of the crossing vortices and antivortices are shifted by $\pi$. This unusual behavior arises from the fact that the six $\bm{\mathit Q}$ vectors are noncoplanar. In contrast, when the condensate $\bm{\mathit Q}$ vectors are on the same plane in the reciprocal space, as is the case with the 3-$\bm{\mathit Q}$ I and 4-$\bm{\mathit Q}$ I states, and those considered in Ref.~\cite{Kamiya2014}, the vortex strings are straight lines along the high-symmetry axis. The helicity of each (anti)vortex increases linearly in the layer index for the 3-$\bm{\mathit Q}$ I state.  In contrast,  the helicity of each (anti)vortex  is shifted  by $\pi$ between consecutive layers of the 4-$\bm{\mathit Q}$ I state. This alternation arises from the fact that $|\bm{\mathit Q}_n| \lesssim 2\pi$ and each [111] layer belongs to a sublattice of the bcc lattice.  In the 6-$\bm{\mathit Q}$ II state, the helicity of each (anti)vortex is constant except for crossing points where it is shifted by $\pi$.

While in this Letter we only focus on the regions where $\omega_{\bm{k}}$ has six degenerate minima, there are other regions in the phase diagrams where vortex crystals should also arise. For example, $\omega_{\bm{k}}$ with eight minima is realized in a region next to the 6-minimum region on both the bcc and fcc lattices, and a 12-minimum case also occurs on the bcc lattice~\cite{supp}. A similar calculation can be applied to these cases in order to obtain the stable multi-$\bm{\mathit Q}$ orderings. 

We note that when some of the effective interactions are attractive (typically the case with ferromagnetic exchanges), the system may undergo a first order phase transition at the saturation field~\cite{Ueda2013,Marmorini2014}. This implies that it is unstable towards  phase separation if one fixes the particle number (i.e., the $z$ component of magnetization) in the canonical ensemble. An alternative scenario is a continuous transition associated with the  condensation of multimagnon bound states~\cite{Ueda2013}. We have verified that none of these cases takes place  for  antiferromagnetic nearest neighbor interactions ($J_1>0$) in both the bcc and fcc lattices. For ferromagnetic nearest neighbor interactions ($J_1<0$), on the other hand, the phase separation occurs in a large region of the phase diagram, as indicated in Figs.~\ref{Fig:phd}(c) and \ref{Fig:phd}(d).

Although crossings of vortex lines have been observed in superconducting vortex glasses and liquids~\cite{Nelson1988,Nelson2002}, we are unaware of the existence of 3D vortex crystals like the one shown in Fig.~\ref{Fig:vortex_string}. These vortex crystals can be detected with neutron diffraction in  single-domain samples. 
Materials with more than two degenerate minima in the magnon dispersion 
of the fully saturated state can be identified directly by measuring the inelastic neutron scattering spectrum at $|H|>|\Hsat|$, or indirectly, by extracting the exchange
constants from the zero-field inelastic neutron scattering spectrum. Nuclear magnetic resonance (NMR) also allows us to distinguish among different multi-$\bm{\mathit Q}$ orderings because 
the NMR line shape is in general qualitatively different for single-, double-, and three-$\bm{\mathit Q}$ orderings.

Our results indicate that these materials are strong candidates to exhibit magnetic vortex 
crystals just below their saturation field.  While here we have considered the particular cases of bcc and fcc lattices as examples, the general principle can be 
directly extended to other highly frustrated structures, such as hcp and pyrochlore lattices, which are also common in nature. Based on our calculations, it is expected that exchange anisotropy (due to e.g., dipolar interactions) will select a double-$\bm{\mathit Q}$ magnetic ordering or a magnetic vortex crystal. The selection mechanism between these two competing phases is provided by the effective interaction between magnons, which ultimately depends on the details of the exchange couplings. There are several candidate materials that comprise highly frustrated 3D lattices of rare-earth magnetic ions. Because the exchange anisotropy is expected to be stronger in these ions due to a large spin-orbit interaction, the vortex crystal phase could extend over a wider window of magnetic field values.

\begin{acknowledgments}
  We would like to thank T.~Momoi and N.~Shannon for helpful discussions.
  Z.W. and A.H.N. were supported by Welch Foundation Grant No. C-1818 and the NSF CAREER award No. DMR-1350237. 
 Z.W. acknowledges support from the CNLS summer student program under which part of the work was performed. 
 A. H. N. was supported by the Cottrell Award from the Research Corporation for Science Advancement (RCSA Grant No. 22799). 
  Work at LANL was performed under the auspices of the
  U.S.\ DOE, Contract No.~DE-AC52-06NA25396, through the LDRD program.
  Y.K. acknowledges financial supports from the RIKEN iTHES project. A.H.N. and C.D.B. thank the Aspen Center for Physics (supported by NSF Grant No. 1066293) for hospitality during the initial stage of this work.
\end{acknowledgments}

\appendix

\setcounter{figure}{0}
\renewcommand{\thefigure}{S\arabic{figure}}
\setcounter{equation}{0}
\renewcommand{\theequation}{S\arabic{equation}}

\newpage
\begin{center}
  {\bf ---Supplemental Material---}
\end{center}

\section{Transformation from the spin to the bosonic language}
\begin{figure}[!htbp]
\includegraphics[width=0.35\textwidth]{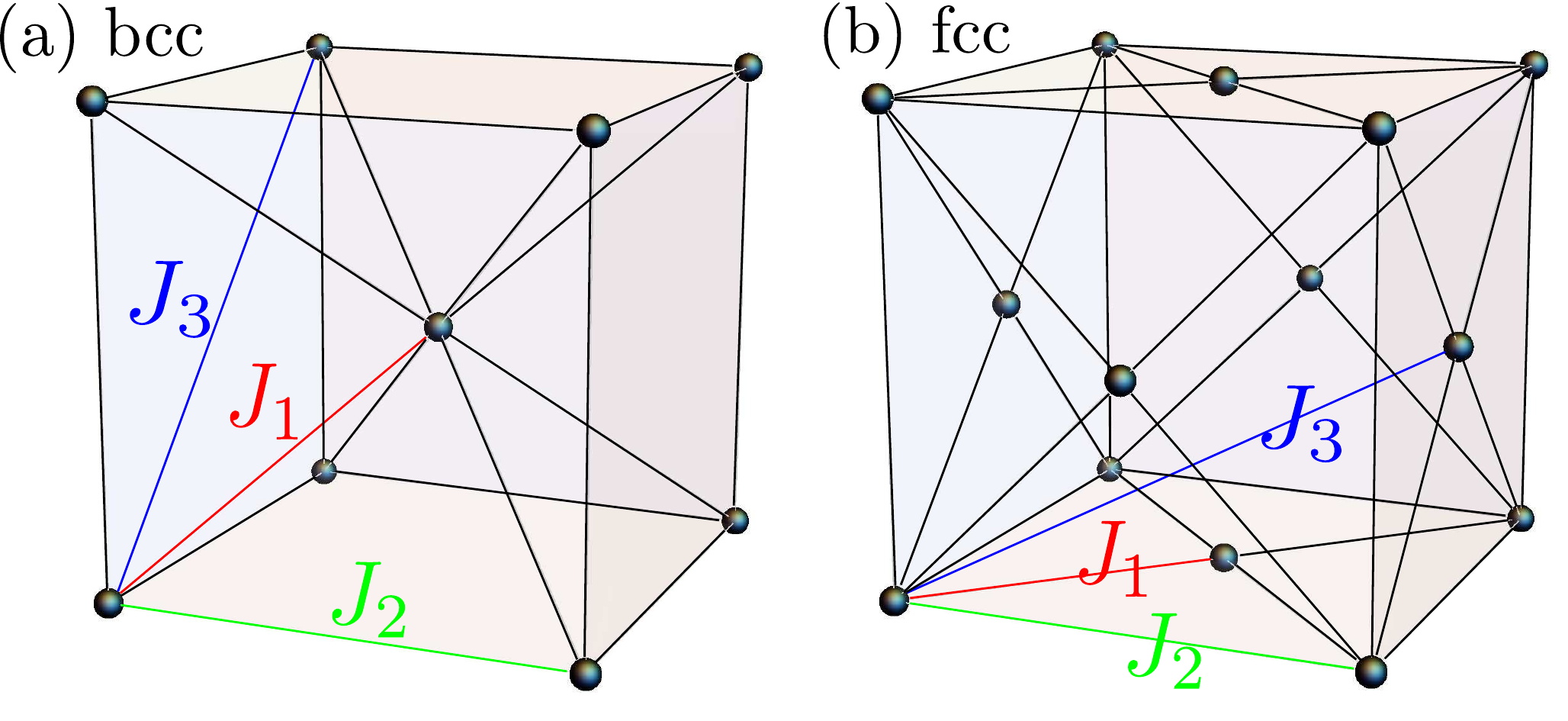}
\caption{The Heisenberg interactions $J_1,J_2,J_3$ are defined on the 1st, 2nd and 3rd nearest neighbors. (a) bcc lattice. (b) fcc lattice.}
\label{Fig:lattices}
\end{figure}

The spin-$\tfrac{1}{2}$ Heisenberg model is defined on the bcc and fcc lattices, shown in Fig.~\ref{Fig:lattices}. By using the Matsubara-Matsuda transformation introduced in the main text, the Hamiltonian is transformed into the bosonic language, up to a constant term:
\begin{align}
 \hat{H}&=\frac{J_1}{2} \sum_{\langle ij \rangle} \left( b_i^\dagger b_j +b_j^\dagger b_i \right) +\frac{J_2}{2} \sum_{\langle \langle ij \rangle \rangle} \left( b_i^\dagger b_j +b_j^\dagger b_i \right)  \nonumber \\
 &\quad +\frac{J_3}{2} \sum_{\langle \langle \langle ij \rangle \rangle \rangle} \left( b_i^\dagger b_j +b_j^\dagger b_i \right)+\frac{U}{2} \sum_i n_i (n_i-1) \nonumber \\
 &\quad  +J_1 \sum_{\langle ij \rangle} n_i n_j+J_2 \sum_{\langle \langle ij \rangle \rangle}n_i n_j +J_3  \sum_{\langle \langle \langle ij \rangle \rangle \rangle} n_i n_j \nonumber \\
 &\quad -\left( \frac{z_1 J_1+z_2 J_2+z_3 J_3}{2}-H \right) \sum_i n_i 
\end{align}
where $U$ is the on-site hard-core repulsion, which is sent to infinity in the calculation, and $z_1,z_2,z_3$ are the coordination numbers of the 1st, 2nd and 3rd nearest neighbors.

By Fourier transformation $b_i^\dagger = \frac{1}{\sqrt{N}} \sum_{\bm{k}} e^{-\iu \bm{k} \cdot \bm{r}_i} b_{\bm{k}}^\dagger$, the Hamiltonian is written down in $\bm{k}$-space:
\begin{align}
 \hat{H}&=\sum_{\bm{k}} \left[ \epsilon (\bm{k})-\epsilon (0)+H \right] b_{\bm{k}}^\dagger b_{\bm{k}} \nonumber \\
 &\quad +\frac{1}{2N} \sum_{\bm{k},\bm{k}^\prime,\bm{q}}(U+V_{\bm{q}}) b_{\bm{k}+\bm{q}}^\dagger b_{\bm{k}^\prime - \bm{q}}^\dagger b_{\bm{k}^\prime} b_{\bm{k}}
\end{align}
where
\begin{equation}
 \epsilon (\bm{k}) \!=\! \frac{J_1}{2} \sum_{\eta_1} e^{\iu \bm{k} \cdot \bm{r}_{\eta_1}}+\!
                      \frac{J_2}{2} \sum_{\eta_2} e^{\iu \bm{k} \cdot \bm{r}_{\eta_2}}+\!
                      \frac{J_3}{2} \sum_{\eta_3} e^{\iu \bm{k} \cdot \bm{r}_{\eta_3}}
\end{equation}
here $\bm{r}_{\eta}$ denote the positions of the neighboring sites. And
\begin{equation}
 V_{\bm{q}}=2 \epsilon (\bm{q})
\end{equation}

To be explicit, For bcc lattice:
\begin{align}
 \epsilon (\bm{k})&=4J_1 \cos \frac{k_x}{2} \cos \frac{k_y}{2} \cos \frac{k_z}{2} +J_2 \Big( \cos k_x \nonumber \\
 &\quad + \cos k_y+ \cos k_z \Big)+2J_3 \Big( \cos k_x \cos k_y \nonumber \\
 &\quad + \cos k_y \cos k_z +\cos k_z \cos k_x \Big)
\end{align}

For fcc lattice:
\begin{align}
 \epsilon (\bm{k})&=2J_1 \Big( \cos \frac{k_x}{2} \cos \frac{k_y}{2}+\cos \frac{k_y}{2} \cos \frac{k_z}{2} \nonumber \\
 &\quad +\cos \frac{k_z}{2} \cos \frac{k_x}{2} \Big)+J_2 \Big( \cos k_x +\cos k_y+ \nonumber \\
 &\quad \cos k_z \Big) +4 J_3 \Big( \cos k_x \cos \frac{k_y}{2} \cos \frac{k_z}{2}+ \nonumber \\
 &\quad \cos k_y \cos \frac{k_z}{2} \cos \frac{k_x}{2}+\cos k_z \cos \frac{k_x}{2} \cos \frac{k_y}{2} \Big)
\end{align}
\begin{figure}[!tbp]
\includegraphics[width=0.45\textwidth]{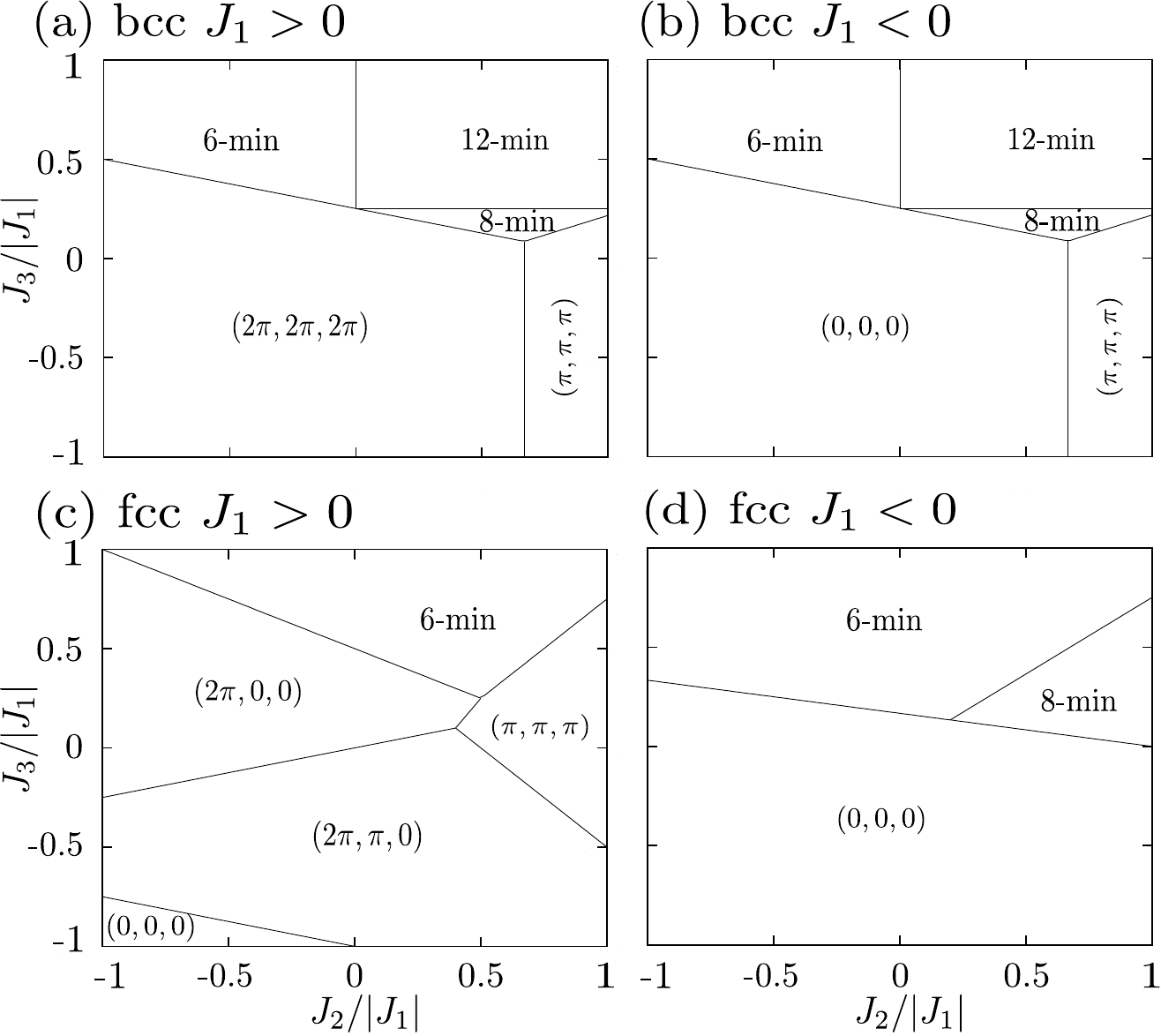}
\caption{The single-magnon phase diagrams, with each region denoted either by the number of minima (when there are multiple minima at incommensurate $\bm{\mathit Q}$-vectors), or denoted by the positions of the $\bm{\mathit Q}$-vectors (when $\bm{\mathit Q}$ is commensurate). (a)(b) bcc lattice. (c)(d) fcc lattice.}
\label{Fig:phd_SingleMagnon}
\end{figure}

We define the minimum value of $\epsilon (\bm{k})$ to be $\epsilon_\text{min}$, in this way $\omega_{\bm{k}}\equiv \epsilon (\bm{k})-\epsilon_\text{min}$ has minimum value equals to zero. The Hamiltonian is rewritten as:
\begin{equation}
\hat{H} \!\!=\!\!\sum_{\bm{k}}\! ( \omega_{\bm{k}}-\mu ) b_{\bm{k}}^\dagger b_{\bm{k}}+\frac{1}{2N} \!\!\sum_{\bm{k},\bm{k}^\prime,\bm{q}} \!\!(U+V_{\bm{q}}) b_{\bm{k}+\bm{q}}^\dagger b_{\bm{k}^\prime - \bm{q}}^\dagger b_{\bm{k}^\prime} b_{\bm{k}}
\end{equation}
where the chemical potential:
\begin{equation}
 \mu=\left[ \epsilon(0)-\epsilon_\text{min} \right]-H\equiv H_\text{sat}-H
\end{equation}

Because of the frustration, the single magnon dispersion $\omega_{\bm{k}}$ can have multiple degenerate minima at different $\bm{\mathit Q}$-vectors. In Fig.~\ref{Fig:phd_SingleMagnon}, we compute the number of minima in $\omega_{\bm{k}}$, for both bcc and fcc lattices.

For concreteness, we focus on the regions with 6 degenerate minima, whose positions are denoted by $\pm \bm{\mathit Q}_n=\pm Q \,{\bf \hat{e}}_n$, where $n=1,2,3$. The value of $Q$ is given by $\cos \tfrac{Q}{2}=-J_1/(J_2+4 J_3)$ for the bcc lattice and $\cos \tfrac{Q}{2}=-(J_1+2J_3)/(J_2+4J_3)$ for the fcc lattice. Correspondingly, the saturation field values are:
\begin{subequations}
\begin{align}
 \Hsat^{\text{bcc}}&=\frac{2 J_1^2}{J_2+4J_3}+4J_1+2J_2+8J_3 \\
 \Hsat^{\text{fcc}}&=\frac{2(J_1+2J_3)^2}{J_2+4J_3}+4J_1+2J_2+16J_3
\end{align}
\end{subequations}

\section{Calculation of Effective Interactions}
\begin{figure}[!htbp]
\includegraphics[width=0.45\textwidth]{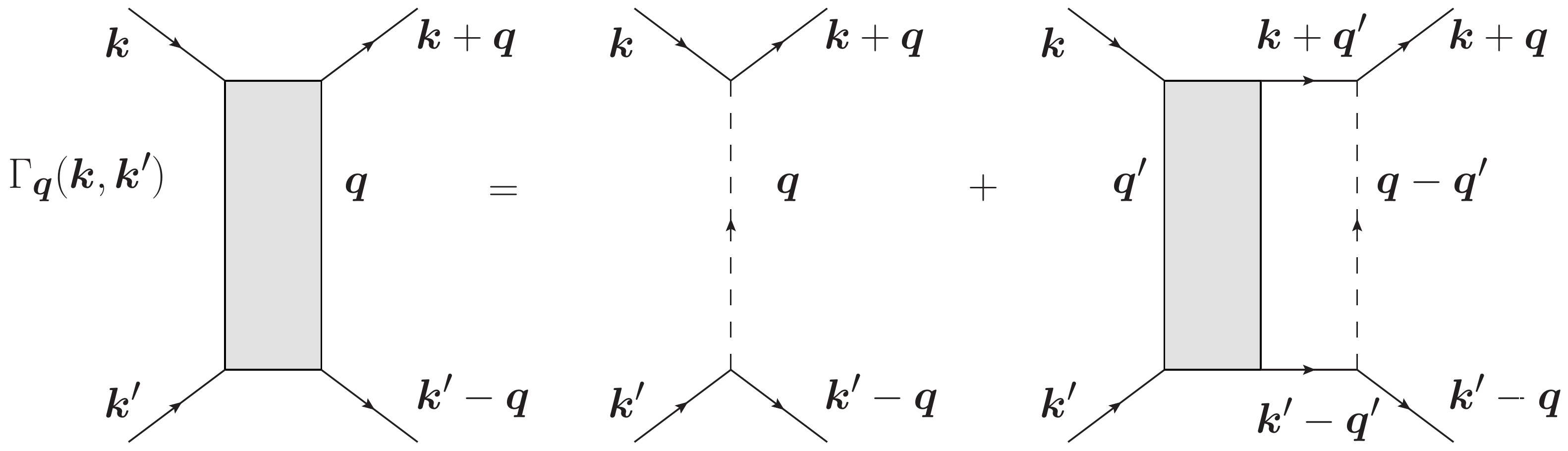}
\caption{Ladder diagrams.}
\label{Fig:ladder}
\end{figure}

The effective interactions in the dilute limit for hard-core bosons are calculated by the Bethe-Salpeter equation, which is equivalent to summing over all the ladder diagrams (Fig.~\ref{Fig:ladder}).
\begin{equation}\label{Eq:Bethe-Salpeter}
\Gamma_{\bm{q}}(\bm{k},\bm{k^\prime})=U+V_{\bm{q}}-\int 
\frac{d^3 \bm{q}^\prime}{V_{\text{BZ}}} \frac{\Gamma_{\bm{q}^\prime}(\bm{k},\bm{k}^\prime)
(U+V_{\bm{q}-\bm{q}^\prime})}{\omega_{\bm{k}+\bm{q}^\prime}+
\omega_{\bm{k}^\prime-\bm{q}^\prime}}
\end{equation}
where $V_{\text{BZ}}$ is the volume of the 1st BZ.

\begin{figure}[!htbp]
\includegraphics[width=0.45\textwidth]{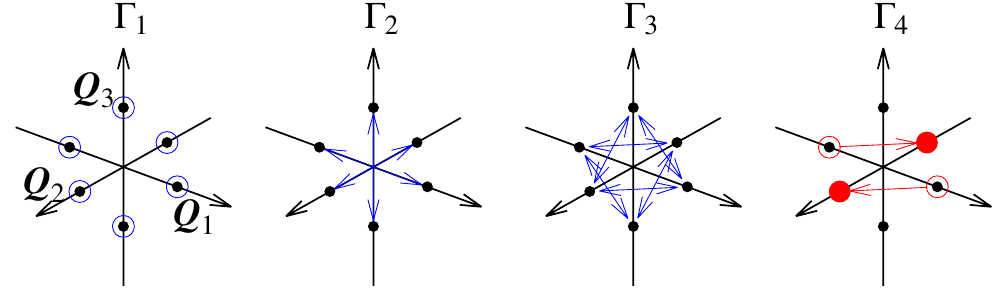}
\caption{Schematic plot of the vertex functions in the long wave length limit.}
\label{Fig:vertex}
\end{figure}

When the magnetic field $H$ is close to the saturation value $H_{\text{sat}}$, 
the system is unstable towards BEC at the dispersion minima. In this 
case we can take the long wave length limit $\bm{k} \rightarrow \pm \bm{\mathit Q}_i$, 
and calculate the corresponding vertex functions (schematically 
shown in Fig.~\ref{Fig:vertex}):
\begin{equation}
\begin{split}
\Gamma_1 &= \Gamma_0 (\bm{\mathit Q}_n, \bm{\mathit Q}_n) \\
\Gamma_2 &= \Gamma_0 (\bm{\mathit Q}_n,-\bm{\mathit Q}_n)+\Gamma_{-2 \bm{\mathit Q}_n} (\bm{\mathit Q}_n,-\bm{\mathit Q}_n) \\
\Gamma_3 &= \Gamma_0 (\bm{\mathit Q}_n,\bm{\mathit Q}_m) +\Gamma_{\bm{\mathit Q}_m-\bm{\mathit Q}_n} (\bm{\mathit Q}_n,\bm{\mathit Q}_m) \\
\Gamma_4 &= \Gamma_{\bm{\mathit Q}_m-\bm{\mathit Q}_n} (\bm{\mathit Q}_n,-\bm{\mathit Q}_n) +
             \Gamma_{-\bm{\mathit Q}_m-\bm{\mathit Q}_n} (\bm{\mathit Q}_n,-\bm{\mathit Q}_n)
\end{split}
\end{equation}

To solve the Bethe-Salpeter equation, we start from the following 
ansatz:
\begin{equation}\label{Eq:ansatz}
\Gamma_{\bm{q}}= \langle \Gamma \rangle
+ \sum_\eta A_\eta V(\bm{r}_\eta) e^{\iu \bm{q} \cdot \bm{r}_\eta}
\end{equation}
where $\bm{r}_\eta$ denotes the positions of the 1st, 2nd, and 3rd neighboring sites. Tthe $\bm{k},\bm{k^\prime}$ index in $\Gamma_{\bm{q}}(\bm{k},\bm{k^\prime})$ 
are omitted for simplicity, and 
$\langle \Gamma \rangle=\int \frac{d^3 \bm{q}^\prime}{V_{\text{BZ}}} \Gamma_{\bm{q}^\prime}$. 
We also assume that $V_{\bm{q}}$ is centro-symmetric, ie
\begin{equation}
\int d^3 \bm{q} V(\bm{q})=0
\end{equation}

By substituting the ansatz into the Bethe-Salpeter equation and taking the 
hard-core limit, we get the following form of linear equations:
\begin{subequations}
\begin{align}
\sum_\eta V(\bm{r}_\eta)(\tau_1^\eta)^* A_\eta +\tau_0 \langle \Gamma \rangle&=1 \\
\sum_\nu (\tau_2^{\eta \nu}V(\bm{r}_\nu)+\delta_{\eta \nu}) A_\nu+ \tau_1^\eta 
\langle \Gamma \rangle &=1
\end{align}
\end{subequations}
where the integrals are defined as:
\begin{subequations}
\begin{align}
\tau_0 &= \int \frac{d^3 q}{V_{\text{BZ}}} \frac{1}{\omega_{\bm{k}+\bm{q}}+
\omega_{\bm{k}^\prime-\bm{q}}} \\
\tau_1^\eta &= \int \frac{d^3 q}{V_{\text{BZ}}} \frac{e^{-i\, \bm{q}\cdot \bm{r}_\eta}}
{\omega_{\bm{k}+\bm{q}}+\omega_{\bm{k}^\prime-\bm{q}}} \\
\tau_2^{\eta \nu} &= \int \frac{d^3 q}{V_{\text{BZ}}} \frac{e^{-i\, \bm{q}\cdot 
(\bm{r}_\eta - \bm{r}_\nu)}}
{\omega_{\bm{k}+\bm{q}}+\omega_{\bm{k}^\prime-\bm{q}}}
\end{align}
\end{subequations}

Denote:
\begin{subequations}
\begin{align}
B_{\eta \nu} &= \tau_2^{\eta \nu} V(\bm{r}_\nu)+\delta_{\eta \nu} \\
C_\eta &= V(\bm{r}_\eta) (\tau_1^\eta)^*
\end{align}
\end{subequations}

The above equations are now organized into a matrix form:
\begin{equation}\label{Eq:mat}
\left(
\begin{array}{cccc}
B_{11} & \cdots & B_{1z} & \tau_1^1 \\
\vdots & \ddots & \vdots & \vdots \\
B_{z1} & \cdots & B_{zz} & \tau_1^z \\
C_1 & \cdots & C_z & \tau_0
\end{array}
\right)
\left(
\begin{array}{c}
A_1 \\
\vdots \\
A_z \\
\langle \Gamma \rangle
\end{array}
\right)
=
\left(
\begin{array}{c}
1 \\
\vdots \\
1 \\
1
\end{array}
\right)
\end{equation}

By solving the linear equations Eq.~\eqref{Eq:mat}, we obtain all the 
unknown coefficients in the ansatz Eq.~\eqref{Eq:ansatz}. Then we can substitute 
the values of $\Gamma_1,\ldots,\Gamma_4$ into the expression of effective 
energy, and determine which multi-$\bm{\mathit Q}$ state will be stabilized.

\section{Effect of symmetric exchange anisotropy}
We consider short-range symmetric exchange anisotropy (cutoff at 2nd nearest neighbor):
\begin{equation}
 \hat{H}_A \propto \sum_{\langle ij \rangle} -3(\bm{S}_i \cdot \bm{r}_{ij})
(\bm{S}_j \cdot \bm{r}_{ij})
\end{equation}
such terms can arise directly from dipole-dipole interactions, or perburbatively from spin-orbit coupling\cite{Moriya1960}.

Similar to the treatment of the Heisenberg exchange interactions, we choose the quantization axis along [111] direction, and represent the spin-$\tfrac{1}{2}$ operators with hard-core bosons. In the long-wavelength limit, for both bcc and fcc lattices:
\begin{align}
 \hat{H}_A & \propto \Big[ 
(\frac{\sqrt{3}}{2}+i\, \frac{1}{2})b_{Q_1}^\dagger b_{-Q_1}^\dagger
+(-\frac{\sqrt{3}}{2}+i\, \frac{1}{2})b_{Q_2}^\dagger b_{-Q_2}^\dagger \nonumber \\
& \quad -i \, b_{Q_3}^\dagger b_{-Q_3}^\dagger  \Big]+h.c
\end{align}

Then we condense the bosons by $\langle b_{\pm \bm{\mathit Q}_n} \rangle/\sqrt{N}= \sqrt{\rho_{\pm \bm{\bm{\mathit Q}_n}}} \exp \left( i \phi_{\pm \bm{\mathit Q}_n}\right)$, which gives the energy correction of symmetric exchange anisotropy:
\begin{equation}
E_A \propto J_A \sum_n \sqrt{\rho_{\bm{\mathit Q}_n} \rho_{-\bm{\mathit Q}_n}} \cos (\Phi_n +2n \pi/3-\pi/2).
\end{equation}
where $\Phi_n=\phi_{Q_n}+\phi_{-Q_n}$.

\bibliography{references}

\end{document}